\documentclass[aps,prd,superscriptaddress,amsmath,amssymb,twocolumn,fleqn,nofootinbib,floatfix]{revtex4-2}
\usepackage[utf8]{inputenc}
\usepackage[hidelinks]{hyperref}
\usepackage{xcolor}
\usepackage{graphicx}
\usepackage{amsmath,amssymb}
\usepackage{bm}
\usepackage{comment}
\usepackage[shortlabels]{enumitem}
\usepackage{overpic}
\usepackage{ulem}
\usepackage{lipsum}
\usepackage{mathtools}
\usepackage{physics}
\usepackage{soul}

\def\Mpc{{\rm Mpc}}

\def\GW{\rm GW}
\def\rhogw{\rho_{\rm GW}}

\def\deltagw{\delta_{\rm GW}}
\def\dpgw{\frac{\delta\rho_{\rm GW}}{\rho_{\rm GW}}}

\newcommand{\vx}{\vec{x}}

\newcommand{\vq}{\vec{q}}
\newcommand{\hc}{\ensuremath{\mathcal{H}}}

\newcommand{\Omegagw}{\ensuremath{\Omega_{\rm GW}}}
\newcommand{\Omegabar}{\ensuremath{\overline{\Omega}_{\rm GW}}}

\newcommand{\del}[2]{\frac{\partial #1}{\partial #2}}

\newcommand{\hn}{\hat n}

\definecolor{greenam}{rgb}{0,0.66,0}

\begin{document}
\title{New universal property of cosmological gravitational wave anisotropies}
\author{Ameek Malhotra}
    \email{ameek.malhotra@unsw.edu.au}
\affiliation{Sydney Consortium for Particle Physics and Cosmology, School of Physics, The University of New South Wales, Sydney NSW 2052, Australia}

\author{Ema Dimastrogiovanni}
    \email{e.dimastrogiovanni@rug.nl}
\affiliation{Van Swinderen Institute for Particle Physics and Gravity, University of Groningen,
Nijenborgh 4, 9747 AG Groningen, The Netherlands}
\affiliation{School of Physics, The University of New South Wales, Sydney NSW 2052, Australia}

\author{\linebreak Guillem Dom\`enech}
    \email{domenech@pd.infn.it}

\affiliation{INFN Sezione di Padova, via Marzolo 8, I-35131 Padova, Italy}
\affiliation{Max-Planck-Institut f{\"u}r Astrophysik, Karl-Schwarzschild-Stra{\ss}e 1, 85748 Garching, Germany}

\author{Matteo Fasiello}
    \email{matteo.fasiello@csic.es}
\affiliation{Instituto de Física Téorica UAM/CSIC, Calle Nicolás Cabrera 13-15, Cantoblanco, 28049, Madrid, Spain}
\affiliation{Institute of Cosmology and Gravitation, University of Portsmouth, PO1 3FX, UK}

\author{Gianmassimo Tasinato}
    \email{g.tasinato@swansea.ac.uk}
    \affiliation{Dipartimento di Fisica e Astronomia, Universit\`a di Bologna, via Irnerio 46, Bologna, Italy}
\affiliation{Physics Department, Swansea University, SA28PP, United Kingdom}
\date{\today}
\begin{abstract}
\noindent
The anisotropies of the stochastic gravitational wave background, as produced in the early phases of cosmological evolution, can act as a key probe of the primordial universe particle content. We point out a new universal property of  gravitational wave anisotropies of cosmological origin: for adiabatic initial conditions, their angular power spectrum is  insensitive to the equation of state of the cosmic fluid
driving the expansion before big-bang nucleosynthesis. Any deviation from this universal behaviour points to the presence of non-adiabatic sources of primordial fluctuations.
Such scenarios can be tested by gravitational wave detectors operating at a frequency range  which is fully complementary to CMB experiments.
In this work we prove this general result, and we illustrate its consequences for a representative realisation of initial conditions based on the curvaton scenario. In the case of the simplest curvaton setup, 
we also find a significant cross-correlation between gravitational wave anisotropies and the CMB temperature fluctuations. There is a fourfold enhancement vis-\`{a}-vis the purely adiabatic scenario. We discuss the implications of our findings for identifying the origin of the (cosmological) gravitational wave background when, as is often the case, this cannot be determined solely on the basis of its spectral shape. 
\end{abstract}

\maketitle

\section{Introduction}
A plethora of early universe
processes are capable of producing a sufficiently large stochastic gravitational wave background (SGWB) to grant detection
via GW experiments (see \cite{Maggiore:1999vm,Caprini:2018mtu} for  reviews). The improved sensitivity of
the next-generation interferometers, such as LISA \cite{amaro2017laser} and ET \cite{Maggiore:2019uih}, may well lead to the detection of such a cosmological SGWB, thus providing us with a new portal into the high-energy phenomena that took place in the primordial universe. 
 
Given the multitude of candidate SGWB sources, it is essential
to fully characterise the stochastic background. The frequency profile is certainly a key observable in identifying the precise the origin of the SGWB (see \cite{Kuroyanagi:2018csn,Caprini:2019pxz} and references therein). However, different processes might produce a SGWB with  similar spectral shapes, thus reducing one's ability to discern among distinct sources.
Primordial gravitational wave non-Gaussianities  do in principle constitute an additional useful handle on the nature of GW sources, but GW propagation effects tend to suppress the size
of  non-Gaussianities  to an unobservably small level    
  \cite{Bartolo:2018rku,Bartolo:2018evs,Margalit:2020sxp} (see \cite{Dimastrogiovanni:2019bfl,Powell:2019kid,Tasinato:2022xyq} for exceptions). 
Crucially, GW anisotropies induced by (ultra) squeezed primordial non-Gaussianity do not suffer from such suppression, and can therefore be of great use in characterising the GW signal \cite{Dimastrogiovanni:2019bfl}.

The origin of GW anisotropies of cosmological nature can be manifold. It may for example be inherent to the SGWB production mechanism~\cite{Kuroyanagi:2016ugi,Jenkins:2018nty,Bartolo:2019yeu,Bartolo:2019oiq,Bartolo:2019zvb,Adshead:2020bji,Malhotra:2020ket,Dimastrogiovanni:2021mfs,Orlando:2022rih}.  GW anisotropies in the early universe have also been studied in the context of GW from phase transitions \cite{Geller:2018mwu,Li:2021iva,Kumar:2021ffi,Bodas:2022zca,Bodas:2022urf}, cosmic strings \cite{Kuroyanagi:2016ugi,Jenkins:2018nty} as well as preheating \cite{Bethke:2013aba,Bethke:2013vca} (see 
\cite{LISACosmologyWorkingGroup:2022kbp} for a recent comprehensive review). Finally, anisotropies may also arise due to GW propagation through an inhomogeneous universe \cite{Alba:2015cms,Contaldi:2016koz,Bartolo:2019yeu,Bartolo:2019oiq}, and it is in this context that we develop the present work.

The recently developed line-of-sight formalism for GW \cite{Contaldi:2016koz,Bartolo:2019yeu,Bartolo:2019oiq} allows one to treat the SGWB anisotropies in the same vein as is done  for the cosmic microwave background (CMB) radiation. Much like the CMB, the SGWB anisotropies can be decomposed into terms that represent respectively the density perturbations at the time of emission, a Sachs-Wolfe (SW) and an integrated Sachs-Wolfe (ISW) effect.\footnote{{Although the total anisotropy is gauge independent, this splitting is not. In this paper we  adopt a Newtonian gauge choice, since it greatly simplifies the analytic calculations.}} This formalism has been recently used to explore the effects that additional relativistic particles and extensions of the $\Lambda$CDM model 
have on SGWB anisotropies and their cross-correlations with the CMB \cite{DallArmi:2020dar,Ricciardone:2021kel,Braglia:2021fxn}.

In this work we point out a universal property of cosmological
SGWB anisotropies: their angular power spectrum is nearly insensitive to the equation of state of the cosmic fluid
driving the universe expansion before big-bang nucleosynthesis (BBN). This result holds under two assumptions. The first  is that the GW initial conditions be set by an adiabatic process. The second requirement is that any transition from a non-standard phase to the standard  radiation dominated (RD) era occurs sufficiently early. 

With the second condition standing, any deviation
from the universal behaviour would point to the presence of non-adiabatic sources of primordial fluctuations. Such fluctuations can therefore  be tested by  probing the SGWB at scales much smaller that than those of the CMB. GW anisotropies thus provide a compelling and complementary handle on the particle content of the very early universe.

Our result on the universality of the GW anisotropies spectrum is significant in that we are able to isolate the mechanism underlying possible deviations: a  departure from adiabaticity in the very early universe.
One interesting example is found in the context of cosmic phase transitions, which can engender a significantly anisotropic SGWB. It was recently shown \cite{Bodas:2022urf} that an  early  phase of  non-standard matter domination  supports sizeable SGWB anisotropies with  significant isocurvature components, compatible with existing constraints from the CMB. 
The properties of the anisotropy spectrum, as the one we discuss in this work, lead to the identification of distinctive and unambiguous signatures  of  non-adiabatic sources for cosmological fluctuations.

The robustness to a non-standard equation of state (when not accompanied by isocurvature modes) that we find in the GW anisotropies signal is quite interesting. Indeed, there exist several cosmological scenarios, well motivated from the top-down perspective,  which are characterised by a different evolution from the standard  RD domination in the early universe expansion (see ref.~\cite{Allahverdi:2020bys} for a review). For example, the coherent oscillations of a scalar field \cite{Turner:1983he} or a period of primordial black hole domination yield a phase of early matter domination. Moreover, in quintessential inflation scenarios \cite{Spokoiny:1993kt,Joyce:1996cp,Ferreira:1997hj,Peebles:1998qn,Joyce:2001hth} there is a period of kinetic energy domination, dubbed ``kination'' after inflation. Note that such kination phase might also occur within the standard radiation era~\cite{Gouttenoire:2021wzu}.  Our results show that, in the absence of isocurvature modes, the impact of such non-standard phases ought to be probed at the level of the SGWB frequency spectrum ~\cite{Giovannini:1998bp,Seto:2003kc,Assadullahi:2009nf,Chung:2010cb,Kuroyanagi:2011fy,Li:2016mmc,Cui:2017ufi,Carr:2017edp,Cui:2018rwi,Inomata:2019ivs,Inomata:2019zqy,DEramo:2019tit,Domenech:2019quo,Bernal:2019lpc,Figueroa:2019paj,Ramberg:2019dgi,Chang:2019mza,Domenech:2020kqm,Domenech:2020ssp,Gouttenoire:2021jhk,Co:2021lkc,Giovannini:2022eue}. This is because under such condition  GW anisotropies are  insensitive to a non-standard evolution.

Given that such universal behaviour is found under the assumption of purely adiabatic sources, it is interesting to consider cases  where the adiabaticity condition does not hold\footnote{This is in line with the intuition expressed in \cite{Alba:2015cms}.}. We do so by focusing on  the curvaton scenario~\cite{Lyth:2001nq,Lyth:2002my}, and identifying the effect of isocurvature fluctuations on GW anisotropies.
We compute explicitly the associated predictions for the angular power spectrum of the anisotropies, highlighting the significant differences with respect to the adiabatic case.
 
\medskip

Our work is organized as follows: We begin with a brief review of the SGWB line-of-sight formalism in Section~\ref{sec:sgwb_los}. In Section~\ref{sec_unirel} we calculate the SGWB anisotropies while taking into account the effects of a non-standard pre-BBN equation of state. Under the assumption of adiabatic initial conditions, we show that the angular power spectrum of the SGWB anisotropies is independent of this non-standard equation of state, leading to a universal prediction for the anisotropies. We will emphasize how the role of the initial condition term, which represents the density perturbation at the time of emission, is crucial to this derivation. 
Isocurvature perturbations are the natural candidate  to break away from the universal behaviour. In Section~\ref{sec_iso}, we focus on a scenario where GW isocurvature perturbations are generated through the curvaton mechanism. We put forward our conclusions in Section~\ref{sec:conclusions}, comment on the implications of these results, and also draw some connections with recent  literature on the SGWB. The appendices contain supplementary details related to the calculations in the main text.

\section{SGWB anisotropies: a line-of-sight formulation}
\label{sec:sgwb_los}
Following \cite{Contaldi:2016koz,Bartolo:2019oiq,Bartolo:2019yeu},
our starting point is the Friedmann-Lemaitre-Robertson-Walker space-time metric,  including scalar perturbations in the Newtonian gauge
\begin{align}
    \label{eq:metric_NG}
    ds^2 = a^2(\eta)\left[-(1+2\Phi)d\eta^2 + (1-2\Psi)d\vec{x}^2\right]\,,
\end{align}
with $a(\eta)$ the scale factor in conformal time, and $\Phi$, $\Psi$ the gravitational
potentials. 
The SGWB can be described by the GW distribution function $f(x^\mu,p^\mu)$,  depending on the GW position $x^\mu$ and  momentum $p^\mu$ (we work in the geometrical optics regime\footnote{{In other words, we consider the propagation of GWs with wavelength much smaller than the current cosmic horizon.}}~\cite{Isaacson_geomoptics,Misner:1973prb}). The total energy density in GW is obtained 
by integrating over momenta: $\rhogw = \int d^3p\,p f(p)$. It is customary to use the spectral energy density parameter $\Omegagw(q)$, defined as \cite{Maggiore:1999vm}
\begin{align}
    \Omegagw = \frac{1}{\rho_{cr}} \frac{d \rhogw}{d\ln q} \; ,
\end{align}
where $q=|p|a$ is the comoving momentum of the gravitons and $\rho_{cr}$ the critical energy density of the universe. The GW  distribution function obeys the following Boltzmann-type equation~\cite{Contaldi:2016koz,Bartolo:2019oiq,Bartolo:2019yeu},
\begin{align}
\label{one-eq}
    \del{f}{\eta} + \del{f}{x^i}n^i+ q\del{f}{q}\left[\del{\Psi}{\eta}-\del{\Phi}{x^i}n^i\right] =0\;.
\end{align}
We can split the homogeneous and isotropic part from an inhomogeneous perturbation, introducing a quantity  $\Gamma$  such that
\begin{align}\label{eq:gammadef}
    f(\vq,\vx)\equiv \bar{f}(q)-\Gamma(\eta,\vec{x},q,\hn)\,\frac{d \,\bar{f}}{d\,{\ln q}}\;.
\end{align}
In Fourier space, the perturbation $\Gamma$ obeys the following linearized equation (primes indicate derivatives w.r.t. conformal time):
\begin{align}
     \Gamma' + ik\mu \Gamma &= \Psi'- ik\mu\Phi,\quad \mu \equiv \hat{k}\cdot\hn,
     \label{eq:Gamma_eq}
\end{align}
with solution \cite{Bartolo:2019oiq,Bartolo:2019yeu},
\begin{align}
\label{solGam}
    \Gamma(\eta_0,k,q,\hn)=& \int_{\eta_i}^{\eta_0} d{\eta}\, \{\delta({\eta}-\eta_i)[{\Phi}(k,\eta)+ \Gamma_I]\nonumber\\
    &+ \Phi'(k,{\eta})+\Psi'(k,{\eta})\}\,e^{- ik\mu(\eta_0-\eta)}\,,
\end{align}
where $\eta_0$ denotes the conformal time today. We denote by $\Gamma_I\equiv \Gamma(\eta_i,k,q)$ the initial condition term, and  with $\delta(\eta-\eta_i)$ the Dirac-delta function over conformal time. The initial condition contribution $\Gamma_I$, first discussed in detail in \cite{Bartolo:2019oiq,Bartolo:2019yeu},
will play an important role in our derivation: we provide more details  on it  in Appendix \ref{app:B}. The anisotropies of the $\Omegagw$, commonly denoted as $\deltagw$, are related to the quantity $\Gamma$ by $\deltagw \equiv [4 - n_\Omega] \Gamma$, with $n_\Omega = \partial\,{\ln \Omegabar(\eta_0,q)}/\partial\,{\ln q}$ parametrising  the tilt of the GW energy density.

Since the anisotropy distribution is a function of the sky location, it is convenient to expand it in spherical harmonics $\Gamma(\hn) = \sum_{\ell m}\Gamma_{\ell m}Y_{\ell m}(\hn)$, and calculate its correlators
\begin{align}
    \langle \Gamma_{\ell m} \Gamma_{\ell'  m'} \rangle \equiv \, C_{\ell}^\Gamma\delta_{\ell \ell'}\delta_{m m'}\,,
\end{align}
under the assumption of statistical isotropy. The spherical harmonic coefficients $\Gamma_{\ell m}$ can be expressed as 
\begin{align}
    \Gamma_{\ell m }= 4\pi (-i)^\ell \int \frac{d^3\vec{k}}{(2\pi)^3}Y^*_{\ell m}(\hat k) T_{\ell}^{\GW}(k),
    \label{eq:Gamma_lm}
\end{align} 

where the function $T^{\GW}_\ell(k)$ combines   the initial condition, the SW, and the ISW terms \cite{Bartolo:2019oiq,Bartolo:2019yeu},

\begin{align}
    T^{\GW}_\ell(k) &= \int_{\eta_i}^{\eta_0} d{\eta}\, \{\delta({\eta}-\eta_i)[{\Phi}(k,\eta)+ \Gamma_I]\nonumber\\
    &+ \Phi'(k,{\eta})+\Psi'(k,{\eta})\}\,j_\ell[k(\eta_0-{\eta})]\}\,.
    \label{eq:def_Tlgw}
\end{align}
We provide in Appendix \ref{app:A} an alternative derivation of the above formula, in terms of the observed graviton energy.

 %%%%%
\section{SGWB anisotropies for adiabatic primordial fluctuations}
\label{sec_unirel}
%%%%%%% 
\label{sec:anisotropies_w}

We  now consider the case of a universe characterised by a non-standard early cosmological  history. Such a possibility is well motivated from  models of high-energy  physics  (see the discussion in the Introduction). Specifically, we  assume
that after inflation, but before
radiation domination (RD), the universe expansion
is driven by a cosmic fluid with an equation-of-state parameter $w_0\neq 1/3$. If GW are generated (or re-enter the horizon) during this phase,\footnote{The corresponding GW  have  frequencies within the reach of  pulsar timing arrays and/or GW interferometers, see e.g. \cite{Maggiore:2018sht}.} they leave distinct  imprints in the frequency profile of the spectrum of $\Omegagw$, as
discussed, for example,  in \cite{Guzzetti:2016mkm,Cui:2017ufi,Domenech:2020kqm,Gouttenoire:2021wzu}.

What is the effect of a $w_0\not=1/3$ on the anisotropies of the SGWB? We now show
that the angular power spectrum for SGWB anisotropies is  {\it  insensitive} to this non-standard phase, as long as  primordial fluctuations are adiabatic and the transition to RD occurs early and rapidly. {One might  expect this result to hold, and for GW anisotropies to closely follow the CMB anisotropies given that the curvature perturbation is conserved on super-horizon scales. We explicitly show why this is the case and how the inclusion of the initial condition term is necessary to erase the effects of any early non-standard expansion history.}  

% \AM{add that this is expected due to conservation of $\zeta$, one diff w.r.t CMB is that changing the eos/expansion rate at photon decoupling does affect the anisotropies however changing eos at GW production does not affect the GW anisotropies.}

Let us prove our claim using cosmological
perturbation theory, by studying the effects
of an early non-standard
cosmology on the anisotropy parameter $\Gamma$ in 
 Eq.~\eqref{solGam}.~We start by noticing that 
on super-Hubble scales the potential $\Phi$ appearing
in Eq.~\eqref{eq:metric_NG} is related to the curvature perturbation in the uniform density gauge $\zeta$ by
\begin{align}
    \Phi = - \frac{3(1+w)}{(5+3w)}\zeta\,\quad;\quad \zeta \equiv -\Psi -\hc\frac{\delta\rho}{\rho'}\,.
    \label{eq:zeta_Phi}
\end{align}
Note that here $\delta\rho$ is the perturbation in the total energy density. One may also define an individual curvature perturbation for each fluid, $\zeta_i$ by replacing $\rho$ for $\rho_i$ in Eq.~\eqref{eq:zeta_Phi}. For adiabatic fluctuations we have $\zeta_i=\zeta$.

An initial equation of state parameter $w_0\neq1/3$ affects the initial value of $\Phi$ through Eq.~\eqref{eq:zeta_Phi}. The subsequent transition to a RD epoch ($w=1/3$) changes the value of the  potential as dictated by
the same equation, resulting in an additional ISW-like effect. The initial condition term  for the GW
evolution appearing in Eq.~\eqref{solGam}
can be computed  with the methods discussed
in \cite{Dimastrogiovanni:2022eir},
and 
 is different from the  radiation domination relation $\Gamma_I=-\Phi/2$. In fact, for a general $w$ and assuming adiabatic primordial perturbations, we find (see Appendix \ref{app:B}) 
\begin{eqnarray}
    \Gamma_I &=& -\frac{2\Phi}{3(1+w)}
    %\\&=& 
    \,=\,\frac{2\,\zeta}{5+3w}\,. \label{eq:Gamma_I}
\end{eqnarray}
 
Here we define adiabatic fluctuations of GWs in standard fashion, with GWs well described by a perfect fluid on cosmological scales. We collect the results obtained so far, and re-evaluate the anisotropy given by Eq.~\eqref{eq:Gamma_lm}. The quantity $T_\ell^{\GW}(k)$ of Eq.~\eqref{eq:def_Tlgw} can be split in two parts as

\begin{align}
   T^{\GW}_\ell(k) =  \int_{\eta_i}^{\eta_r}\ldots+\int_{\eta_r}^{\eta_0}\ldots\,.
   \label{eq:TGWspl}
\end{align}
Here  $\eta_r$ the conformal time at the transition from the early $w_0\neq 1/3$ epoch to the standard RD era and the dots refer to the integrand in Eq.~\eqref{eq:def_Tlgw}. 
The second term on the right-hand side of Eq.~\eqref{eq:TGWspl} corresponds to an ISW effect associated with the standard $\Lambda$CDM universe, and is common to all scenarios irrespective of the initial equation of state. We calculate this term using \texttt{CAMB}~\cite{Lewis:1999bs} assuming the \textit{Planck} bestfit values for the $\Lambda$CDM parameters \cite{Planck:2018vyg}. 
\begin{figure}
    \centering
    \includegraphics[width=0.45\textwidth]{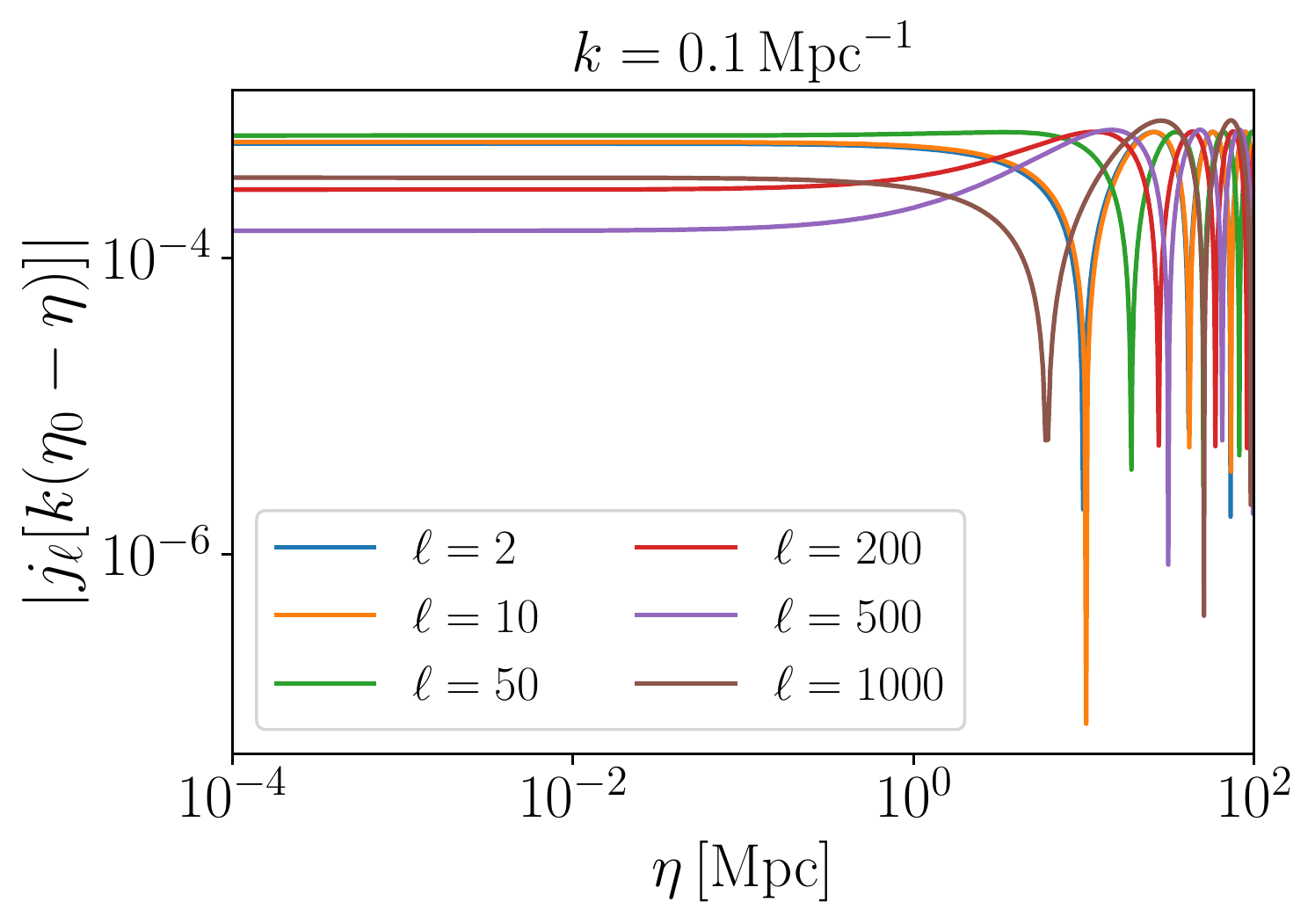}
    \caption{A graphical demonstration of the validity of the approximation involving the spherical Bessel functions, as used in Eqs.~\eqref{eq:sourcegw_int} and \eqref{eq:Tl_by_jl1}. We selected 
   $k=0.1\,\Mpc^{-1}$ as an example. For a given $k$, the Bessels are essentially constant as long as $k\eta\ll 1$, roughly corresponding to the duration over which the mode $k$ remains super-Hubble.}
    \label{fig:bessel}
\end{figure}

Notice that $\eta_0\gg \eta_r,\eta_i$.
For instance, in units of \mbox{$c=1$}, and considering a transition redshift \mbox{$z>10^8$}, $\eta_r$ is  of the order \mbox{$\eta_r\lesssim 10^{-4}$} whereas $\eta_0 \sim 10^4$. Thus, for the large scale modes of interest, we  always have \mbox{$k\eta \ll 1$} when \mbox{$\eta_i<\eta <\eta_r $}.
We can then approximate $k(\eta_0-\eta)\simeq k\eta_0$ in the argument of the spherical Bessel functions in the first integral of Eq.~(\ref{eq:TGWspl}). See also Fig.~\ref{fig:bessel}, which graphically supports
this approximation. As a result, the first term in the right-hand side of Eq.~\eqref{eq:TGWspl} can be approximated by
\begin{align}
    \label{eq:sourcegw_int}
    T^{\GW (1)}_\ell(k) &\approx\big(\Phi(\eta_i) +\Gamma_I(\eta_i)\nonumber\\&+\left[\Phi(k,\eta)+\Psi(k,\eta)\right]^f_i\big)j_\ell[k \eta_0]\nonumber\\&+O(\eta_r/\eta_0)\,.
\end{align}
Note that this result holds even if there are intermediate phases between eras with equations of state $w = w_0$ and  $w=1/3$.

Since after the transition to standard expansion we have a radiation dominated universe, Eq.~\eqref{eq:sourcegw_int}  reads
\begin{align}
   \frac{T^{\GW(1)}_\ell(k)}{j_\ell[k\eta_0]}\approx&\,\Gamma_I(w_0)-\Psi(w_0)\nonumber\\&+\Phi(1/3)+\Psi(1/3)\,.
     \label{eq:Tl_by_jl1}
\end{align}
The combination of the first two terms in Eq.~\eqref{eq:Tl_by_jl1} corresponds to the definition of the curvature perturbation associated to gravitational waves, namely
\begin{align}
\Gamma_I-\Psi=-\Psi+\frac{1}{4}\frac{\delta\rho_{\rm GW}}{\rho_{\rm GW}}\equiv\zeta_{\rm GW}\,.
\end{align}
It follows that, since the initial conditions are set on superhorizon scales by a constant $\zeta$, Eq.~\eqref{eq:Tl_by_jl1} is insensitive to the early equation of state of the universe. We can also check this explicitly using Eq.~\eqref{eq:zeta_Phi}, which yields
\begin{align}
    \label{eq:gamma_minus_psi}
    \Gamma_I(w_0)-\Psi(w_0) &= -\Phi\left[1+\frac{2}{3+3w_0}\right] 
    \,= \,\zeta,
\end{align}
assuming  no anisotropic stress. This relation is valid at the early times, so that $\Phi=\Psi$. Since $\Phi(1/3)=-2\zeta/3$, we conclude that
\begin{align}
    \frac{T^{\GW(1)}_\ell(k)}{j_\ell[k(\eta_0)]} &= -\frac{4}{3}\zeta+\zeta= -\frac{1}{3}\zeta,
\end{align}
irrespective  of the equation of state $w_0$. The quantity $T^{\GW}_\ell$ defined in Eq.~\eqref{eq:def_Tlgw} can then be written as

\begin{align}
    T^{\GW}_\ell = & \int_{\eta_r}^{\eta_0}d\eta\,[\Phi(k,\eta)'+\Psi(k,\eta)']j_\ell[k(\eta_0-\eta)]\nonumber\\
    &-\frac{1}{3}\zeta\, j_\ell[k\eta_0].
    \label{eq:Tlgw_ad}
\end{align}
Thus, the dependence on the equation of state parameter $\omega_0$, associated with in the initial phase
of expansion, 
 has completely disappeared from the final result.

\begin{figure}
    \centering
    \includegraphics[width=0.45\textwidth]{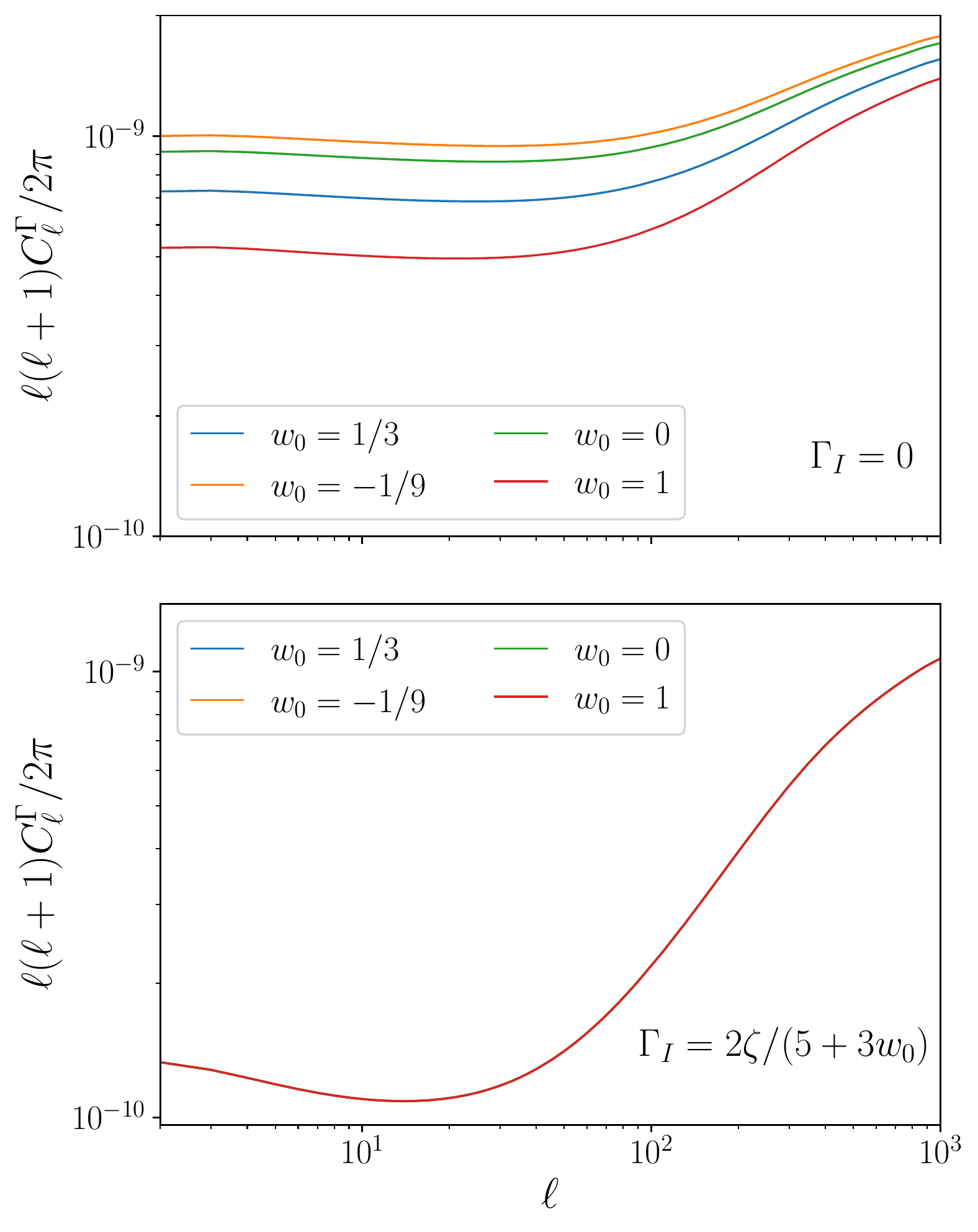}
    \caption{
    The upper plot shows the effect of varying $w_0$ on the SGWB anisotropies without including the initial condition term. The lower plot includes the contribution from the initial condition contribution $\Gamma_I$.}
    \label{fig:Cl_Gamma_w}
\end{figure}

It is important
to stress the essential role of the initial condition
contribution \eqref{eq:Gamma_I} for our arguments, derived
under the assumption of adiabaticity of the primordial fluctuations. The angular power spectrum of the GW anisotropy is shown in Fig.~\ref{fig:Cl_Gamma_w} for several values of $w_0$, corroborating our conclusions. It is   the presence of the initial condition contribution in Eq.~\eqref{solGam} which removes any effect of the non-standard equation of state.

\medskip

The  approximation involving the spherical Bessel functions, used in Eqs.~\eqref{eq:sourcegw_int} and \eqref{eq:Tl_by_jl1}, can be intuitively understood as follows. In real space, the ISW effect involves an integral along the GW geodesic, hence  it is sensitive to the (temporal and spatial) variation of the potential along the GW line-of-sight. However, during the early transition from non-standard to RD cosmology, GW only cover an infinitesimal comoving distance, with $k\Delta\eta\ll 1$. Thus, the spatial gradients of the potentials ($\sim k\Delta\eta\times \Phi$) can be neglected. Moving to Fourier space, this implies that the spherical Bessel functions appearing in the first integral of Eq.~\eqref{eq:TGWspl} can be approximated by a constant. Small changes in the time of emission -- and hence in the corresponding equation of state --  do not leave any imprints in the anisotropies of the SGWB.  

This result holds for GW generated ``early-enough'', such that the large-scale modes of interest are still super-Hubble: adiabaticity ensures the conservation of the curvature perturbation independently of any changes in the equation of state. Additionally, the non-standard cosmic  phase needs to occur very early on in the cosmic history for our arguments to hold. This is not so stringent as an assumption, since the universe must be radiation dominated already by the time of BBN\footnote{See \cite{Kawasaki:1999na,Kawasaki:2000en,Hasegawa:2019jsa} for 
lower bounds of O(MeV) on the reheating temperature from BBN constraints.}, which itself happens early (at $z\sim 10^8$ or equivalently $T\sim 100\,\mathrm{keV}$). %\textcolor{cyan}{For instance, the lower bound on the reheating temperature is roughly $T>4 \,{\rm MeV}$.}
In the end what is important is the equation of state when the long wavelength mode re-enters the horizon: as long as that is not affected by the early phase, the anisotropy spectrum remains unchanged. {For the largest observable scales relevant for gravitational anisotropies, this happens during radiation or matter domination.}

{Our result also highlights the importance of properly accounting for the initial condition term $\Gamma_I$, one may otherwise end up with a spurious dependence on the initial equation of state (as seen in Fig.~\ref{fig:Cl_Gamma_w}). Let us also briefly comment regarding the initial time $\eta_i$ in Eq.~\eqref{eq:Tlgw_ad}. In general, this should be taken to be the time when the GW are produced/emitted (in the CMB case, this corresponds to the time of photon decoupling). The derivation in this section shows that for a given long wavelength mode $k$, we may also take $\eta_i$ to be the around the time when the mode re-enters the horizon, i.e. when $\zeta_k$ starts evolving. In the adiabatic case, both choices lead to the same result and there is no dependence on any initial non-standard equation of state. A point of difference from the CMB is that changes in the equation of state at recombination would affect the CMB anisotropies, especially on intermediate and small angular scales. On the other hand, GW anisotropies are completely unaffected by changes in the equation of state at the time of emission.} 

\medskip

Interestingly, our results also have consequences 
for early universe phenomena involving Standard Model
physics only, e.g. quantum chromodynamics
(QCD) phase transition. In fact, during the QCD phase transition, which occurs at temperatures
$T\sim 100$ MeV,  the equation of state of the universe  changes: this fact has interesting implications for primordial black hole formation~\cite{Byrnes:2018clq,Franciolini:2022tfm,Escriva:2022bwe} as well for the SGWB, see e.g. \cite{Saikawa:2018rcs,Abe:2020sqb}. But, as demonstrated above, it does not affect the SGWB anisotropies ($C_\ell^\Gamma$) for adiabatic primordial perturbations. The frequency dependence of the observed $C_\ell^{\GW}$ is still sensitive to the effects of the QCD phase transition through the $\Omegagw$ spectrum since $C_\ell^{\GW} = (4-n_\Omega)^2 C_\ell^\Gamma$. However, such an effect in the GW anisotropy  spectrum does not provide any additional information, with respect to what we can learn from the frequency profile of $\Omegagw$.

\section{SGWB anisotropies with isocurvature contributions}
\label{sec_iso}
Given our robust predictions  for the universal properties
of SGWB anisotropies from adiabatic initial conditions, it is  interesting  
 to explore  possible  consequences of abandoning  the adiabaticity  assumption. We do so in this Section, studying the effects of primordial isocurvature GW perturbations for the SGWB anisotropies in concrete early universe
 scenarios. Our aim in this Section is {\it not} to specifically examine the consequences of non-adiabatic initial conditions for a SGWB in the context of
 a non-standard equation of state. Instead, we wish to investigate  the implications of isocurvature  initial conditions for SGWB anisotropies, showing explicitly that they change
 the universal predictions we derived
 in the previous Section for the adiabatic case. 

\medskip

We start by reviewing the definition of GW isocurvature fluctuations, and discuss their consequences on GW anisotropies.  First, let us consider  a case in which  the GW initial conditions are set during the radiation era. Isocurvature fluctuations depend on the difference between contributions to  the curvature fluctuation
from different species. 
 In particular, 
a GW isocurvature component, when defined with respect to the standard model radiation bath, can be expressed as \cite{Kumar:2021ffi},
\begin{align}
    S_{{\GW}, r} = 3(\zeta_{\GW}-\zeta_r),\quad  \zeta_x = -\Psi - \hc\frac{\delta\rho_x}{{{\rho}}_x'},
    \label{eq:S_GW_zeta_i}
\end{align}
where $x=\{{\GW},\,r\}$.  

 Eq~\eqref{eq:S_GW_zeta_i} 
leads to the relation~\cite{Kumar:2021ffi}
\begin{align}
    \Gamma_I = \zeta_{\GW}+\Psi = \zeta_r+\frac{1}{3}S_{{\GW},r}-\frac{2}{3}\zeta\,,
\end{align}
where, in RD,  $\Psi=-2\zeta/3$.
We can now relate the value of $\zeta_r$ to the curvature perturbation in terms of the total energy density, assuming that 
the universe contains only radiation and the GW background. Namely,
\begin{align}
    \zeta &= -\Psi -\hc\frac{\delta\rho}{\rho'}= \zeta_r + \frac{1}{3}f_{\GW}S_{{\GW},r}\,.
\end{align}
We introduce the quantity $f_{\GW}$  defined as
\begin{align}\label{eq:fgw}
    f_{\GW} = \frac{(1+w_{\GW})\rhogw}{\sum_{x}(1+w_x)\rho_x}.
\end{align}
with $w_x$ the equation of state parameter of the component~$x$. 
Thus, one obtains as final result\footnote{A similar calculation for CMB temperature fluctuations yields $\Theta_I=\zeta/3-f_{\rm GW}S_{\rm GW,r}/3$, where $\Theta_I$ are the initial temperature fluctuations. Thus, the effects of such isocurvature perturbations on the CMB anisotropies are suppressed by a factor $f_{\GW}\ll 1$ w.r.t their effects on the GW anisotropies \cite{Kumar:2021ffi}.}  \cite{Kumar:2021ffi}
\begin{align}
    \Gamma_I = \frac{1}{4}\dpgw = \frac{\zeta}{3} + \frac{1}{3}(1-f_{\GW})S_{{\GW},r}.
    \label{eq:gamma_I_iso}
\end{align}
We can further develop this line of reasoning, and generalize these findings to scenarios in which the SGWB is produced during an epoch when  the background energy density is dominated by a component $x$ with an arbitrary equation of state $w_0$ (not necessarily radiation), as done in the previous Section. In this case, since $f_{\GW}\ll 1$, $\zeta\approx \zeta_{x}$, we can use Eq.~\eqref{eq:zeta_Phi}, 
define the GW isocurvature  with respect to a fluid with equation of state $\omega_x$, 
and obtain
\begin{align}
    \Gamma_I \simeq \frac{2\zeta_x}{5+3w_0} + \frac{1}{3}S_{\GW,x},\quad S_{\GW,x}=3(\zeta_{\GW}-\zeta_{x}).\label{eq:Gamma_I_iso}
\end{align}
generalizing Eq.~\eqref{eq:gamma_I_iso}. Notice that  
Eq.~(\ref{eq:Gamma_I_iso}) differs from \eqref{eq:Gamma_I}, due to the contribution of the isocurvature perturbations.  This term affects
the arguments of the previous  Section, and  can lead to significant departures from the standard adiabatic result of Eq.~\eqref{eq:Tlgw_ad} for the anisotropy angular correlations.

\medskip

We now concretely investigate this possibility by building explicit scenarios leading to isocurvature contributions, with the aim of analyzing  their consequences for the angular correlations of GW anisotropies. Refs.~\cite{Kumar:2021ffi,Bodas:2022zca} have previously considered cosmological models producing isocurvature GW perturbations
 from phase transitions during RD. We  develop an alternative 
 perspective for generating   GW isocurvature perturbations from inflation, based on the
 curvaton mechanism.

\subsection*{Curvaton scenario\label{subsec:curvaton}}

The curvaton model~\cite{Enqvist:2001zp,Moroi:2002rd,Lyth:2001nq,Lyth:2002my} posits that during inflation, besides the inflaton, a spectator field is present,  in the form of  a subdominant scalar field $\chi$. This field  is essentially massless, and is characterized by  a non-vanishing vacuum expectation  value $\chi_*$. The curvaton  fluctuations $\delta\chi$, as developed during inflation, are initially isocurvature. As cosmic expansion proceeds, at some epoch during the post-inflationary evolution, the curvaton mass overcomes the Hubble friction, and $\chi$ undergoes coherent oscillations about the minimum of its potential, behaving like dust.  At this stage, the curvaton can constitute the dominant contribution to the energy budget of the universe, with its initial isocurvature fluctuations  converted into curvature fluctuations. After this epoch of curvaton dominance,  we assume that the curvaton decays to Standard Model particles. For our purposes, in order to derive analytical results,  we focus on the case of instantaneous curvaton decay.  This process
 can affect the SGWB and CMB anisotropies, to a degree that depends on the energy budget of the curvaton at the time of its decay.

\begin{figure}
    \centering
    \includegraphics[width=0.4\textwidth]{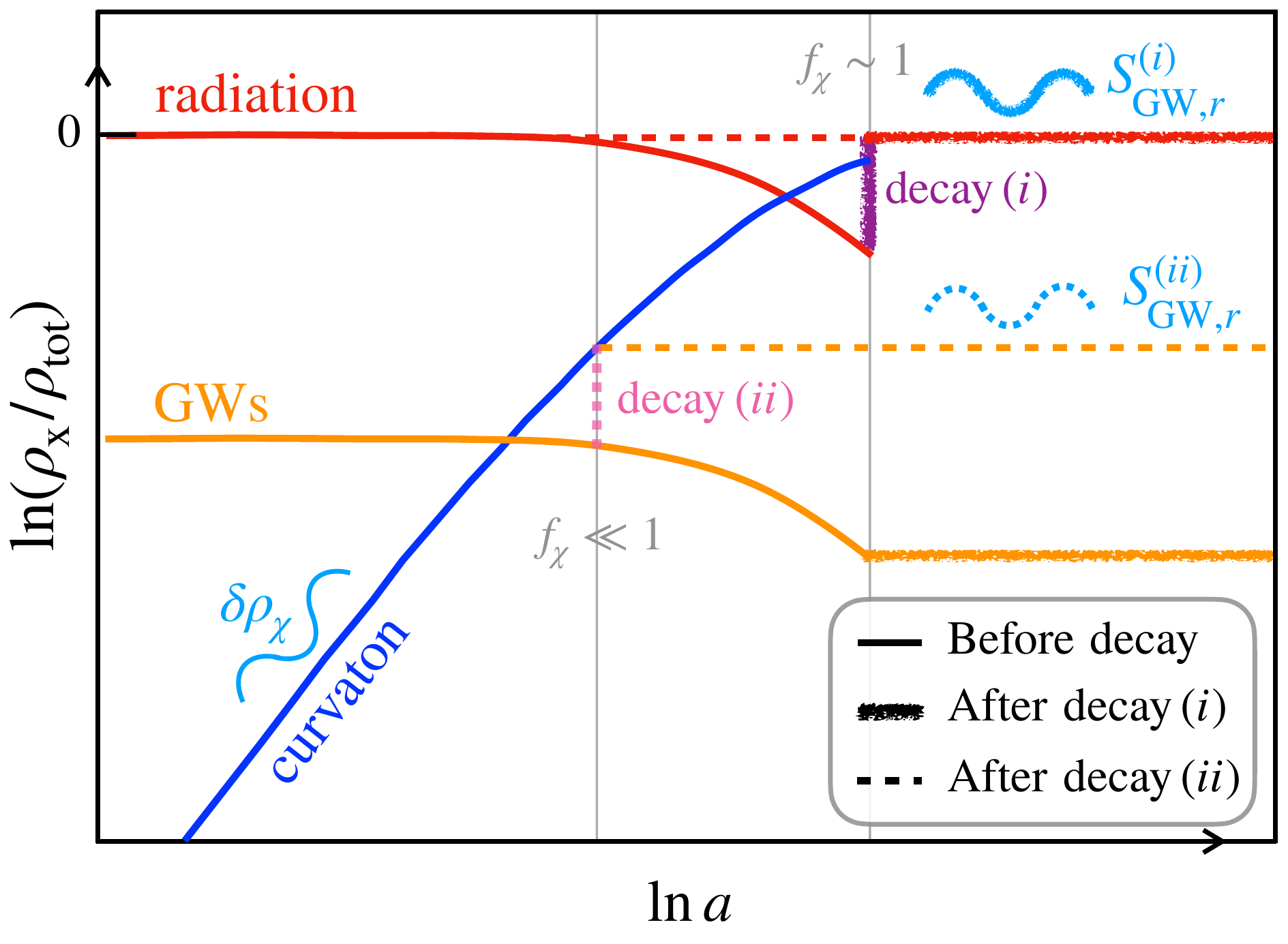}
    \caption{Illustration of the curvaton mechanism and its implications for initial GWB isocurvature fluctuations. We show the logarithm of energy density of a given fluid (standard radiation r, GWs and the curvaton $\chi$) normalised to the total energy density as a function of e-folds or $\ln a$. Note that we assume $w_\chi=0$ and we take arbitrary initial background densities for illustrative purposes. At some point, the curvaton decays either to standard radiation (case $(i)$) or to GWs (case $(ii)$). The fraction of the curvaton at the time of decay for $(i)$ is $f_\chi\sim 1$ while for $(ii)$ is $f_\chi\ll1$. Then the initial isocurvature fluctuations due to the curvaton are either transferred to standard radiation in case $(i)$ or to GWs in case $(ii)$. Due to the asymmetric decay of the curvaton, there remains an isocurvature component between radiation and GWs, labelled $S_{\GW,r}$.}
    \label{fig:curvaton}
\end{figure}

We envision 
two possible mechanisms (pictorially 
represented in Fig.~\ref{fig:curvaton}) 
for generating GW isocurvature perturbations
through a 
 curvaton field: 

\begin{enumerate}[label={\textit{(\roman*)}}]
    \item The curvature perturbation originates from an isocurvature-to-adiabatic conversion of primordial fluctuations,  after the   curvaton decays. Gravitational waves, on the other hand, are generated during inflation, or during another early universe phase well before the curvaton dominates. The curvaton decays into Standard Model particles, and therefore its initial isocurvature component only survives within the SGWB fluctuations.
    \item Gravitational waves are generated through the dynamics of the curvaton itself --  e.g. small scale curvaton fluctuations  source GWs as in~\cite{Bartolo:2007vp}. The anisotropies in the GW energy density are then correlated with the curvaton fluctuations at the largest cosmological scales. However, in contrast to case~$(i)$,  we assume  that the curvaton energy density remains subdominant  until its decay. It follows that the curvaton contribution to the total curvature perturbation is negligible. This set-up shares some similarities with the dynamics of the isocurvature mode in
    scenarios including the effects of dark radiation~\cite{Ghosh:2021axu}, and  phase-transitions~\cite{Bodas:2022zca}. 
\end{enumerate}

Let us now proceed to concretely analyze these configurations (see refs.~\cite{Lyth:2004gb,Sasaki:2006kq,Bartolo:2002vf,Byrnes:2014xua} for studies on CMB fluctuations when the curvaton mechanism  is in place). We shall assume that, after inflation, the universe  contains three species of fluids. Their energy densities are denoted by $\rho_{x}$, and their equations of state via $w_{x}$, with $x=\{r,\chi,\rm GW\}$. The first fluid $\rho_r$, corresponding to radiation, dominates the universe immediately after inflation.\footnote{One can take a more general approach, and consider an arbitrary equation of state after inflation. This possibility does not qualitatively change our results, hence we do not pursue it any further.} The second fluid is the curvaton, with energy density $\rho_\chi$: as explained above, this field  decays at some epoch after inflation. The third fluid corresponds to the GW energy density, $\rho_{\rm GW}$, which can be treated as a subdominant (GW) radiation component throughout the cosmic evolution. 
Each component is characterized by an associated curvature perturbation given by
\begin{align}\label{eq:definition}
\zeta_x=-\Psi +\frac{\delta\rho_x}{3(1+w_x)\rho_x}\,.
\end{align}
The curvature perturbation $\zeta$ on uniform density slices is given by Eq.~\eqref{eq:zeta_Phi}.
The isocurvature fluctuation, as defined in terms of two distinct components $x$ and $y$, is defined similarly as above, as  
\begin{align}
S_{x,y}=3(\zeta_x-\zeta_y)\,.
\end{align}
The expressions for the quantities $\zeta_x$ and $S_{x,y}$ are gauge independent. Thanks to this property, we are  then free to evaluate the initial conditions in a uniform curvature slicing, finding 
\begin{align}
\zeta_{\chi,\rm ini}=\frac{1}{3(1+w_\chi)}\left(\frac{\delta\rho_\chi}{\rho_\chi}\right)_*\,.
\label{eq:zeta_chi_density}
\end{align}
The subscript $^*$ means that we evaluate the quantities   at horizon crossing during inflation. For the sake of generality, we do not fix a specific  equation of state for the curvaton $\chi$ contribution.\footnote{In general, $\delta\rho_\chi$ is an arbitrary function of $\chi$; e.g. for a potential $m_\chi^2\chi^2$, one finds $\delta\rho_\chi/\rho_\chi= 2\delta\chi/\chi_*$.} We assume that the inflaton decays into  radiation, implying that  $\zeta_{r,\rm ini}$ coincides with the curvature fluctuation generated during the inflationary process.

Since all components individually obey an energy conservation condition,  and are not characterized by non-adiabatic pressure, each of the three curvature perturbations $\zeta_x$ are individually conserved 
during cosmic evolution \cite{Lyth:2004gb,Sasaki:2006kq}, except at the time of curvaton decay. Assuming an instantaneous curvaton decay, we compute $\zeta$ right \textsl{before} and \textsl{after} the decay, which we respectively denote by $\zeta^b_{\rm dec}$ and $\zeta^a_{\rm dec}$. We assume a uniform density slicing ($\sum_x\delta\rho_x=0$). Using \mbox{$\delta\rho_x /\rho_x =3(1+w_x)(\zeta_x-\zeta)$}, we find 
\begin{align}\label{eq:zetadecay_before}
\zeta^b_{\rm dec}=f^b_\chi \zeta_{\chi,\rm ini}&+f^b_{\rm GW} \zeta_{\rm GW, ini}\nonumber\\&+(1-f^b_{\chi}-f_{\rm GW}^b)\, \zeta_{r, \rm ini}\,.
\end{align}
The quantity $f_\chi$ is defined as $f_{\GW}$ in Eq.~\eqref{eq:fgw} but replacing the subscript $\GW$ for $\chi$. The notation before (indicated with the superscript $b$) and after (superscript $a$) is important, since the curvaton can decay into standard radiation, and/or gravitational waves. In fact, the fraction $f^a_{\rm GW}$ can be different than $f^b_{\rm GW}$, depending on how much GW energy is generated in  the  decay process of $\chi$.

Hereafter, for concreteness we assume $f^b_{\rm GW}\ll1$ and neglect its contribution. 
For simplicity, we  also assume that there is no initial isocurvature fluctuation between radiation and gravitational waves, so $\zeta_{r,\rm ini}=\zeta_{\text{GW,ini}}$ (unless otherwise stated). Notice that the curvature perturbation \textit{after} the decay is given by
\begin{align}\label{eq:zetadecay}
\zeta^a_{\rm dec}=f^a_{\rm GW} \zeta^a_{\rm GW, dec}+(1-f_{\rm GW}^a)\, \zeta^a_{r, \rm dec}\,.
\end{align}
In Appendix \ref{app:curvaton}
we present further details on the evolution of this system. 

Isocurvature fluctuations are constant on superhorizon scales, except at the time of curvaton decay. In fact, the resulting isocurvature depends on the end products of the curvaton annihilation. In what follows, we  perform two separate studies of scenarios $(i)$ and $(ii)$. 

\subsubsection*{Case (i)}
\label{sec:iso_a}
In this case, the curvaton field decays into   radiation. In combination with the conservation condition for $\zeta$ across the instantaneous decay, this implies that \mbox{$\zeta^a_{\rm r, dec}\approx \zeta^a_{\rm dec}\approx \zeta^b_{\rm dec}$}. The resulting isocurvature contribution 
$S_{{\GW},\rm r}$

after the curvaton decays is 
\begin{align}
S^{a%,{(i)}
}_{{\GW},\rm r | dec}&\equiv 3(\zeta^a_{\rm GW,dec}-\zeta^a_{\rm r,dec})\nonumber\\&\approx 3(\zeta_{\rm GW,ini}-\zeta^b_{\rm dec})\approx f_\chi^b S_{{\GW},\chi |  ini }\,,
\end{align}
where we use Eq.~\eqref{eq:zetadecay_before}. We learn that the initial curvaton  isocurvature fluctuation is inherited by the GW background, but  with a suppression factor $f_\chi$. This is because a fraction $f_\chi$ of the total radiation is made out of the decay of the curvaton $\chi$, which  is characterized by an initial isocurvature fluctuation $S_{{\GW},\chi}$ with respect to GW.

To better appreciate the consequences of these isocurvature contributions for SGWB anisotropies, we  focus on an explicit, simple example. We set  initial conditions $\zeta_{\chi,\rm ini}\gg \zeta_{r,\rm ini}=\zeta_{\GW,\rm ini}$ (the equality assumes initial adiabatic GW fluctuations after inflation). Then, at the time of curvaton decay to radiation, we have  
\begin{align}
    \zeta^a_{r,\rm dec} \simeq f^b_\chi\zeta_{\chi,\rm ini}\,.
\end{align}
In this example, the contribution of radiation
to the curvature fluctuation after curvaton decay,
$\zeta^a_r$, is also responsible for sourcing  CMB fluctuations. For this reason, the value of the amplitude  $f^b_\chi\zeta_{\chi,\rm ini}$ is fixed by observations. The curvaton equation of state enters this amplitude via Eq.~\eqref{eq:zeta_chi_density}, although its effect is degenerate with those of $f_\chi$ and $\delta\rho_\chi/\rho_\chi$. A measurement of the anisotropies would then constrain the combination \mbox{$f_\chi (1+3w_\chi)^{-1} \delta\rho_\chi/\rho_\chi $}.

We focus on SGWB modes re-entering the horizon during radiation domination for which we use Eq.~\eqref{eq:gamma_I_iso} and $f_{\GW}\ll 1$ to obtain
\begin{align}
    \Gamma_I &\simeq \frac{1}{3}\zeta^a_{r,\rm dec} + \frac{1}{3}S^a_{{\GW},r} = \frac{1}{3}\zeta^a_{r,\rm dec} + \frac{1}{3}f^b_\chi S_{ \GW \chi,\text{ini}} \nonumber\\&\simeq -\frac{2}{3}f^b_\chi \zeta_{\chi,\rm ini}=-\frac{2}{3}\zeta^a_{r,\rm dec}\,,\label{eq:Gamma_I_curv_a}
\end{align}
where $\Gamma_I$ are the GW fluctuations after the decay of the curvaton and the start of standard Big Bang cosmology.
Note that in this case the initial condition term is significantly different from the adiabatic case of Eq.~\eqref{eq:Gamma_I} where $\Gamma_{\rm ad}=\zeta/3$. Now, evaluating the total anisotropy with these modified initial conditions using Eq.~\eqref{eq:def_Tlgw} and the subsequent results of Section~\ref{sec_unirel}, we obtain
\begin{align}
    T^{\GW}_\ell =& -\frac{4}{3}\zeta^a_{r,\rm dec}\times j_{\ell}[k\eta_0] \nonumber\\ 
    &+ \int_{\eta_r}^{\eta_0}[\Phi(k,\eta)'+\Psi(k,\eta)']j_\ell[k(\eta_0-\eta)].
    \label{eq:Tlgw_iso_a}
\end{align}
  
%\st{We learn that the first term on the right-hand side  is 4 times the adiabatic result of Eq.~\eqref{eq:Tlgw_ad} while the second term is the same.} 
The SGWB map is then also  completely correlated with the CMB in this case since both are sourced by the initial fluctuations of $\chi$. 
{This cross-correlation is given by
\begin{align}
   \langle \Gamma_{\ell m} \Delta_{\ell' m'}^{T(E)} \rangle \equiv C_\ell^{\Gamma \,T(E)}\delta_{\ell \ell'}\delta_{m m'}
\end{align}
where $\Delta_{\ell m}^{T(E)}$ denotes the spherical harmonic coefficients of the CMB temperature or $E$-mode polarisation anisotropies.
{We also see that the first term on the right-hand side of Eq.~\eqref{eq:Tlgw_iso_a} is 4 times the adiabatic result of Eq.~\eqref{eq:Tlgw_ad} while the second term is the same. Since the first term which is a SW-like term, dominates on large angular scales, an SGWB$\cross$CMB correlation which, on the largest scales is \textit{4 times larger} compared to the standard adiabatic result would strongly hint towards the simplest curvaton scenario. One can also understand this from Fig.~\ref{fig:Cl_iso} by noticing that the larger SW term in this case leads to nearly flat spectrum for the $C_\ell^\Gamma$, in contrast to the adiabatic case where the ISW bump is visible at large-$\ell$ (see also~\cite{DallArmi:2020dar}).}

\subsubsection*{Case (ii)}
\label{sec:iso_b}

We proceed with the scenario $(ii)$, where only the curvaton sources GWs while its energy density remains sub-dominant. In this case, if all of the curvaton  energy density goes into GWs, we have that $\delta\rho_{\rm GW,\rm dec}\approx\delta\rho_{\chi,\rm dec}$. We also assume $\zeta_{\chi,\rm ini}\gg \zeta_{r,\rm ini}$, but $f_\chi^b\zeta_{\chi,\rm ini}\ll \zeta_{r,\rm ini}$, which is the most interesting phenomenological case.  With these assumptions we arrive at (see Appendix \ref{app:curvaton})
\begin{align}
\label{35}
S^{a,(ii)}_{{\GW}, r|\rm dec}&\equiv 3(\zeta^a_{\rm GW, dec}-\zeta^a_{r,\rm dec})\approx 3(\zeta^a_{\rm GW, dec}-\zeta^b_{r,\rm dec})\nonumber\\&\approx3\frac{(1+w_\chi)}{(1+w_r)}\zeta_{\chi,\rm ini}\,,
\end{align}
where we use the following relations, valid in a uniform density slicing, assuming $\zeta^a_{\rm dec}=\zeta^b_{\rm dec}\ll\zeta_{\chi,\rm ini}$:
\begin{align}
\zeta^a_{\rm GW, dec}&=\zeta^a_{\rm dec}+\frac{1}{3(1+w_r)}\frac{\delta\rho^a_{\GW}}{\rho^a_{\GW}}\nonumber\\&\approx \frac{1}{3(1+w_r)}\frac{\delta\rho^b_{\chi}}{\rho^b_{\chi}}\approx \frac{(1+w_\chi)}{(1+w_r)}\zeta_{\chi,\rm ini}\,.
\end{align}
Essentially all the GW isocurvature component originates from the large curvaton  fluctuations, hence we  need to take into account the change in equation of state. Notice that the energy density of the curvaton is subdominant, and therefore it does  not significantly  affect the total curvature perturbation.

In general, not all of the curvaton energy density is  transferred to GWs. For GWs generated by means of the curvaton decay, it follows that
\begin{align}
    T^{\GW}_\ell \approx&\,\left[\frac{(1+w_\chi)}{(1+w_r)}\zeta_{\chi,\rm ini} -\frac{1}{3}\zeta^a_{r,\rm dec}\right]j_{\ell}[k\eta_0] \nonumber\\&+ \int_{\eta_r}^{\eta_0}[\Phi'(k,\eta)+\Psi'(k,\eta)]
    j_\ell[k(\eta_0-\eta)]\,,
    \label{eq:Tlgw_iso_b}
\end{align}
where we used that $f^b_\chi\zeta_{\chi,\rm ini}\ll \zeta_{r,\rm ini}$, hence  CMB fluctuations are set entirely by $\zeta_r$. In case $(ii)$ one can have large isocurvature fluctuations, i.e. $\zeta_{\chi,\rm ini}\gg \zeta^a_{r,\rm dec}$, while having a small impact on the CMB since \mbox{$f^b_\chi \ll1$}. In this case, the correlation between the CMB and SGWB anisotropies is much smaller, in contradistinction to case~$(i)$.\footnote{SGWB and CMB anisotropies could instead have a larger correlation if $f_{\chi}^b\ll1$ and, contrary to what we assumed just above Eq.~(\ref{35}), we have $\zeta^b_\chi\ll \zeta^b_r$. Under these assumptions one finds, using Eq.~(\ref{eq:sgw2}), that $S_{\GW, r}\approx -{9}(1+w_\chi) \zeta_r/4$. SGWB anisotropies are then larger by approximately a factor 3 with respect to the CMB for $w_\chi=0$.}

Importantly, a large amount of GW isocurvature requires $f^b_\chi\ll1$, which can lead to large non-Gaussianities in the SGWB anisotropies, as it would happen in the standard curvaton scenario \cite{Sasaki:2006kq}. This since as $f^b_\chi$ decreases, the expectation value $\chi_*$ should also decrease (for a fixed curvaton mass); as $\zeta_\chi$ is large, it implies that higher order terms in $\delta\chi/\chi_*$ become more relevant. We leave a detailed study of this scenario for future work. 
Such non-Gaussian signatures are also expected for scenarios similar in spirit  to this case $(ii)$, i.e. GWs generated by subdominant fields with large isocurvature fluctuations. This has  already been  pointed out in \cite{Bodas:2022zca}.

\subsubsection*{A summarizing plot}
In Fig.~\ref{fig:Cl_iso} we plot the angular power spectrum of the anisotropies for cases $(i)$ and $(ii)$, as given by Eqs.~\eqref{eq:Tlgw_iso_a} and \eqref{eq:Tlgw_iso_b} respectively. 
We clearly notice that the presence of the isocurvature perturbation leads to a strong departure from the adiabatic relation of Fig.~\ref{fig:Cl_Gamma_w}. In both cases, the much larger isocurvature component significantly enhances the SGWB anisotropies relative to the adiabatic case. For the same reason, the spectrum is essentially flat across all scales, similar to the large scale SW plateau in the CMB. Moreover, the amplitude and tilt for case $(i)$ is fixed but for case $(ii)$ it is not. In the latter case, the $\ell$-dependence of $C_\ell^\Gamma$  depends crucially on the spectral shape of $\mathcal{P}_{\zeta_\chi}(k)$ (see \cite{Bodas:2022zca} for an example). The plot also includes a more phenomenological set-up where $S_{{\GW},r}=3\zeta_r$, which corresponds to an isocurvature component equal in magnitude but opposite in sign compared to case $(i)$.  In this scenario, the isocurvature is anti-correlated with the GW, leading to a reduction in power on large scales compared to $(i)$ and, for the same reason, to an anti-correlation between the  SGWB and CMB maps. One could realise such anticorrelation if, for example, the initial isocurvature fluctuations of the curvaton are already anti-correlated with initial adiabatic fluctuations. This is possible within general two-field models of inflation \cite{Amendola:2001ni,Bartolo:2001rt,Byrnes:2006fr}, whose dynamics is different from the simplest curvaton scenario. We leave this for future work.

\begin{figure}
    \centering
    \includegraphics[width=0.45\textwidth]{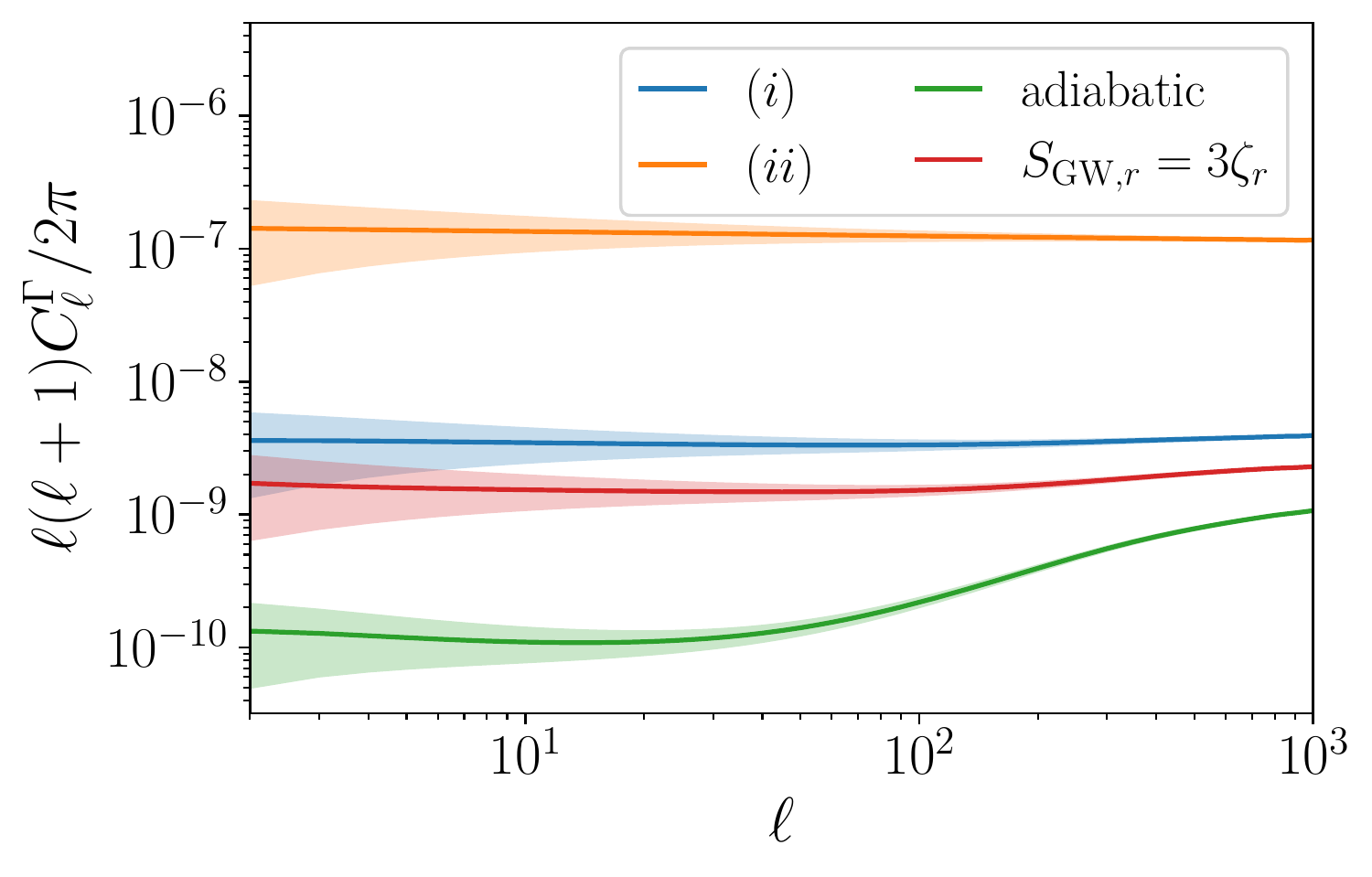}
    \caption{The angular power spectrum of the SGWB anisotropies for cases $(i)$ and $(ii)$. For case $(i)$ the isocurvature amplitude is fixed (see Eq.~\eqref{eq:Tlgw_iso_a}). For case $(ii)$ we have chosen $|\zeta_\chi|=10|\zeta_r|$ and $w_\chi=0$. The adiabatic prediction and a scenario with the sign of $S_{{\GW},r}$ opposite to that of case $(i)$ are also shown for comparison. The quantity $\zeta_r$ is determined by the CMB amplitude $\mathcal{P}_{\zeta_r} = 2.09\times 10^{-9}$ and the spectral tilt $n_s=0.9649$. The shaded regions denote the cosmic variance limited error bars~\cite{Tegmark:1996bz}.}
    \label{fig:Cl_iso}
\end{figure}

\medskip

\section{Discussion and conclusions}
\label{sec:conclusions}

There exist a wide variety of gravitational wave production mechanisms in the early universe. The ever-growing interest in such possibilities relies on the discovery potential associated with the detection of GW of cosmological origin. From learning the energy scales at which the ``cosmological collider'' operates  (e.g. during inflation), to the possibility of testing  beyond-the-Standard-Model physics (e.g. via first order phase transitions), from key clues on pre-heating dynamics to important lessons on cosmic strings and possibly dark matter,  a great deal of progress in our understanding of the early universe will result from the detection and characterisation of a primordial GW signal.

It is then crucial to develop a most effective toolbox aimed at identifying (1) the astrophysical \textit{vs}  cosmological nature and (2) the precise origin of a given GW stochastic background. The study of the spectral shape, chirality, and non-Gaussianity of the SGWB is certainly part of the standard ``characterisation algorithm''. Our focus in this work has been on another critical property of the spectrum: the presence of an anisotropic component.    

GW anisotropies provide an additional handle on inflationary models and interactions, on the presence of large scale inhomogeneities in the early universe, and so on. The central question we set out to address has been on the possibility of testing the equation of state (EoS) of the early universe through its effect on anisotropies. 
A changing  EoS is motivated for example at the QCD phase transition. More in general, a non-standard EoS may result from the coherent oscillations of a scalar field during a period of kination, and several other well-motivated scenarios. 

Interestingly, we find that, under specific assumptions, a universal behaviour is in place: GW anisotropies are insensitive to the EoS of the early universe. This robustness of the anisotropies profile to deviations from a standard evolution history holds if the transition to radiation domination occurs sufficiently early and provided that primordial fluctuations are adiabatic. The fact that these are relatively mild assumptions underscores the wide range of validity of the universal behaviour we uncovered.

Conversely, deviations from the universal formula point clearly to the presence of isocurvature fluctuations in the early evolution of the universe. We exploited this notion in two specific realisations of the well-known curvaton scenario, obtaining in case~$(i)$ up to a four-fold enhancement   (w.r.t. the adiabatic case) of the GW anisotropies due to the presence of the isocurvature fluctuations. This amounts to over an order of magnitude increase in terms of the anisotropies angular power spectrum. The fact that the effects of the isocurvature perturbations are significant on large angular scales is also remarkable in that such scenarios may be tested in the future~\cite{Cui:2022ani}, despite the limited angular resolution of GW detectors~\cite{Baker:2019ync,Alonso:2020rar,Mentasti:2023icu}.

Whenever the curvaton comes to give a significant contribution to the curvature perturbation, and if the leading GW are generated independently during inflation,
we found (case $(i)$) that cross-correlations of SGWB anisotropies with those of the CMB can also be used as an extremely effective probe of the curvaton hypothesis. This provides an additional instrument in our curvaton diagnostics that is complementary to, for example, CMB constraints on the non-linear parameter $f_{\rm NL}$.

It will be important to further explore  
deviations from the universal condition we identified here in several directions, going well beyond the (simplest) curvaton scenario. We plan to study the 
effects of isocurvature modes on GW anisotropies in a variety of interesting early universe setups and present our findings in future work.

\section*{Acknowledgments} 
G.D. would like to thank Misao Sasaki for useful discussions. We would also like to thank Sabino Matarrese for helpful insights on the initial conditions of GW fluctuations. G.D. as a Fellini fellow was supported by the European Union's Horizon 2020 research and  innovation programme under the Marie Sk{\l}odowska-Curie grant agreement No 754496.~M.F. would like to acknowledge support from the ``Atracci\'{o}n de Talento” grant 2019-T1/TIC15784, his work is
partially supported by the Agencia Estatal de Investigaci\'{o}n through the Grant IFT Centro de Excelencia Severo Ochoa No CEX2020-001007-S, funded by MCIN/AEI/10.13039/501100011033. GT is partially funded by the STFC grant ST/T000813/1. For the purpose of open access, the authors have applied a Creative Commons Attribution licence to any Author Accepted Manuscript version arising.
	
\appendix

\section{Heuristic derivation of GWB anisotropies}
\label{app:A}

In this Appendix we provide a simpler, less rigorous, derivation of the GWB anisotropies, which is equivalent to the collisionless Boltzmann formulation. We follow the analogy of CMB anisotropies given in Sec.~2.5 of ref.~\cite{Durrer:2020fza}.

Consider that we receive from a direction $n^i$ a collection of (massless) gravitons with energy $E=-k^\mu u_\mu$ which were emitted in the early universe and that propagated through a perturbed FLRW universe. $k^\mu$ is the 4-momentum of the graviton which follows null-geodesics and $u_\mu$ is the observer's velocity. If we compare the energy of the emitted graviton with the received one in the Newton (shear-free) gauge we have that \cite{Durrer:2020fza}
\begin{align}
\frac{E_{\rm obs}}{E_{\rm emit}}&=\frac{a_{\rm emit}}{a_{\rm obs}}\left(1-\frac{\delta q}{q}\Big|_{\rm emit}-\delta(k^\mu u_\mu)^{\rm obs}_{\rm emit}\right)\nonumber\\&=\frac{1}{1+z}\left(1-\frac{\delta q}{q}\Big|_{\rm emit}-\frac{\delta z}{1+z}\right)\,,
\end{align}
where we assumed initial energy fluctuations $\delta q$ at the surface of emission and we defined
\begin{align}\label{eq:deltaz}
\frac{\delta z}{1+z}=\left[V_in^i+\Phi\right]^{\rm obs}_{\rm emit}-\int^{\rm obs}_{\rm emit} d\lambda\left(\Psi'+\Phi'\right)\,.
\end{align}
In Eq.~\eqref{eq:deltaz}, $\lambda$ is the affine parameter of the null geodesics and $V_i$ is the 3-velocity of the fluid.

We now relate the initial graviton energy fluctuations to the Boltzmann formalism of the main text. We note that we do not detect single gravitons but a distribution of energy density, represented by the distribution function $f(x,q)$. The energy density of the GW background is then given by
\begin{align}\label{eq:rhogwsdef}
\rho_{\rm GW}(x,\eta)=a^{-4}(\eta)\int d^3q \,q f(x,q,\eta)\,,
\end{align}
where we set today's scale factor to $a_0=1$. Note that at the background level we have that $f=\bar f(q)$ and the time dependence in \eqref{eq:rhogwsdef} only enters through the scale factor and thus satisfies energy conservation, i.e. $\bar\rho_{\rm GW}'+4\hc\bar\rho_{\rm GW}=0$. From Eq.~\eqref{eq:rhogwsdef} we see that any small initial inhomogeneity in the distribution function, say $f(x,q,\eta_i)=\bar f (q,\eta_i)+\delta f(x,q,\eta_i)$, can be thought of an inhomogeneous distribution of graviton momentum as
\begin{align}
f(x,q,\eta_i)=\bar f (q+\delta q,\eta_i)=\bar f (q)+\frac{\partial \bar f}{\partial q}\delta q\,.
\end{align}
Comparing with the definition of $\Gamma$ from Eq.~\eqref{eq:gammadef}, that is $\delta f=-q\tfrac{\partial \bar f}{\partial q}\Gamma_I$ we identify
\begin{align}
\frac{\delta q}{q}\Big|_{\rm emit}\equiv-\Gamma_I\,.
\end{align}
With this result and Eq.~\eqref{eq:deltaz} we arrive at the conclusion that the observed graviton's energy anisotropies are given by
\begin{align}
\Gamma=\delta\left(\frac{E_{\rm obs}}{E_{\rm emit}}\right)=\Gamma_I&-\left[V_in^i+\Phi\right]^{\rm obs}_{\rm emit}\nonumber\\&+\int^{\rm obs}_{\rm emit} d\lambda\left(\Psi'+\Phi'\right)\,.
\end{align}
This is exactly the same as Eq.~\eqref{solGam} if one neglects the direction independent monopole at the location of the observer, the dipole due to our motion and use that on superhorizon scales the initial velocities are suppressed by a factor $k^2/\hc^2$ and so are negligible. The last step is to use the fact that we do not detect graviton's energies but the spectral density of the GWB, which yields
\begin{align}
\delta_{\rm GW}=\frac{\delta \Omega_{\rm GW}}{\Omega_{\rm GW}}=\frac{q^4\delta f}{\Omega_{\rm GW}}=(4-n_\Omega)\Gamma\,
\end{align}
where we used that $\rho_{\rm GW}=3H^2M_{\rm pl}^2\int d\ln q\, \Omega_{\rm GW}$.

\section{Derivation of initial condition term}
\label{app:B}

The initial condition $\Gamma_I$ is a model-dependent term that represents the perturbation to the GW distribution function at the time of emission/production. Its monopole $\Gamma_I^{(0)}=\int d^2\hn\,\Gamma_I/4\pi$ represents the initial GW density perturbation and is the counterpart of the CMB quantity $\Theta_0$, the monopole of the photon density (or equivalently temperature) fluctuation at recombination~\cite{dodelson2020modern}. 

We now derive the contribution to the GW initial condition term that arises from adiabatic primordial perturbations. In our analysis, we neglect any higher order terms and take $\Gamma_I=\Gamma_I^{(0)}$ since the large scale modes of interest are super-Hubble at the initial time, suppressing these higher order terms.  

One can use the 00-component of the perturbed Einstein's equations in the Newtonian gauge to get \cite{dodelson2020modern},
\begin{align}
    3{\cal H}^2 \Phi= -4\pi G a^2 \rho \delta \implies \delta = -2\Phi
\end{align}
where $\delta$ denotes the density contrast for the dominant component of the universe. Then, by adiabaticity
\begin{align}
    \frac{\delta\rhogw}{(1+w_{\GW})\rhogw} = \frac{\delta}{(1+w)} = -\frac{8\Phi}{3(1+w)} 
\end{align}
and finally using the results of ref.~\cite{Dimastrogiovanni:2022eir}
\begin{align}
   \frac{\delta\rhogw}{\rhogw} = 4\Gamma_I,
\end{align}
which holds in this case since $\Gamma_I$ is independent of the GW frequency. Note that the exact time when the initial conditions should be set for GWs is either at GW generation (if GWs are generated by sub-horizon processes) or some time after horizon re-entry.

Alternatively, one can generalise the method presented in sec 2.1.1 of \cite{Dimastrogiovanni:2022eir} to arbitrary $w$ and obtain the same result (see also \cite{Creminelli:2011sq} for the original application to the CMB).

\section{General formulas for the curvaton GW isocurvature}
\label{app:curvaton}

Here we present the exact formulas for the GW isocurvature after curvaton decoupling without assuming any type of initial conditions. We consider case $(i)$ and $(ii)$ separately first and then we provide the general formula. 

First, for case $(i)$ we have that the curvaton only decays to radiation and, therefore, by continuity we have that after the curvaton decays $\rho^a_r=\rho^b_r+\rho^b_\chi$ and $\delta\rho^a_r=\delta\rho^b_r+\delta\rho^b_\chi$. The notation $b$ and $a$ respectively refers to evaluation just before and after the curvaton decays. With these relations, one can find that
\begin{align}
   S^{a,(i)}_{\rm GW r}&=3(\zeta^a_{\rm GW,dec}-\zeta^a_{r,\rm dec})\nonumber\\&=-3\frac{\rho^b_\chi+\rho^b_r+\rho^b_{\rm GW}}{\rho^b_\chi+\rho^b_{\rm r}}\Big[(1-f^b_{\GW}-f^b_{\chi})\zeta_{r,\rm ini}\nonumber\\&+f^b_{\chi}\zeta_{\chi, \rm ini}-(1-f^b_{\GW})\zeta_{\rm GW, ini}\Big]\,.
\end{align}
In deriving this equation we made use of the definition of the curvature perturbation \eqref{eq:definition}. In the main text we studied the case $f^b_{\GW}\ll1$, which leads to $S^{a,(i)}_{\rm GW r, dec}\approx-3f^b_\chi(\zeta_{\chi,\rm ini}-\zeta_{r,\rm ini})$. Note that if all curvature perturbations are equal then isocurvature vanishes as it should.

We proceed similarly for case $(ii)$, using that the curvaton now decays only to GWs, that is $\rho^a_{\GW}=\rho^b_{\GW}+\rho^b_\chi$ and $\delta\rho^a_{\GW}=\delta\rho^b_{\GW}+\delta\rho^b_\chi$. Then, we obtain
\begin{align}\label{eq:sgw2}
    S^{a,(ii)}_{\rm GW r,dec}&=3(\zeta^a_{\rm GW,dec}-\zeta^a_{r,\rm dec})\nonumber\\&=3\frac{\rho^b_\chi+\rho^b_r+\rho^b_{\rm GW}}{\rho^b_\chi+\rho^b_{\rm GW}}\Big[f^b_{\chi}\zeta_{\rm \chi,\rm ini}\nonumber\\&+f^b_{\GW}\zeta_{\rm GW,ini}-(f^b_\chi+f^b_{\GW})\zeta_{r,\rm ini}\Big]
\end{align}
In the case when GWs are mainly sourced by the curvaton, so that $f^b_{\rm GW}\ll1$, and the curvaton has large fluctuations $\zeta_\chi\gg\zeta_r$ (but $f^b_\chi\zeta_{\chi,\rm ini}\ll\zeta_{r,\rm ini}$ because $\rho^b_\chi\ll\rho^b_r$), we find that $S^{a,(ii)}_{\rm GW r,dec}\approx3\frac{(1+w_\chi)}{(1+w_r)}\zeta_{\chi,\rm ini}$.

In the most general case where only a fraction $\sigma$ of the curvaton energy decays into GWs, i.e. $\rho_{\GW}^a=\rho^b_{\GW}+\sigma\rho^b_\chi$, we find that
\begin{align}
    S^a_{\rm GW r,dec}&=3(\zeta^a_{\rm GW,dec}-\zeta^a_{r,\rm dec})\nonumber\\&=3\frac{(\rho^b_\chi+\rho^b_r+\rho^b_{\rm GW})^2}{(\rho^b_{\rm GW}+\sigma\rho^b_\chi)(\rho^b_{\rm r}+(1-\sigma)\rho^b_\chi)}\nonumber\\&\times\Big[(\sigma\omega^b_r-(1-\sigma)\omega^b_{\GW})f^b_{\chi}\zeta_{\chi,\rm ini}\nonumber\\&+(\omega^b_r+\frac{1+w_\chi}{1+w_r}(1-\sigma)\omega^b_{\chi})f^b_{\GW}\zeta_{\rm GW,ini}\nonumber\\&-(\omega^b_{\GW}+\frac{1+w_\chi}{1+w_r}\sigma\omega^b_\chi)f^b_{\rm r}\zeta_{r, \rm ini}\Big]\,,
\end{align}
where we have defined
\begin{align}
\omega_x\equiv\frac{\rho_x}{\rho_\chi+\rho_r+\rho_{\rm GW}}\Big|_{\rm dec}\,.
\end{align}
It is straightforward to check that we recover case $(i)$ when $\sigma\to0$ and case $(ii)$ when $\sigma\to 1$. We also checked that such formula for GW isocurvature vanishes for adiabatic initial conditions.

\section{Scalar induced GWs and the SGWB spectral shape} 
\label{app:C}

In this appendix we provide an example of a scenario where the same spectral shape can be generated via different production mechanisms. Our example shall be that of a peaked broken power law spectral shape, which can arise in SGWB from first order phase transitions \cite{Caprini:2019egz}, kination~\cite{Gouttenoire:2021wzu}, cosmic domain walls~\cite{Saikawa:2017hiv} and scalar induced GW. Importantly, even if the first three mechanisms produce SGWB with distinguishable spectral shapes, i.e. different power law indices on either side of the peak, we will demonstrate here that for each of the three mechanisms, one can produce the same spectral shape with scalar induced GW.

The induced GW spectrum is approximately a broken power-law with a peak in two cases:~(a) the primordial spectrum is a broken power-law and~(b)~the equation of state of the primordial universe is negative~\cite{Domenech:2019quo,Domenech:2020kqm,Atal:2021jyo,Balaji:2022dbi}. To illustrate our point regarding the degeneracy in the spectral shape, it suffices to focus only on case~(a).  In case (a), if the primordial spectrum around the peak scale $k_{pk}$ is given by
\begin{align}
    P_{\zeta} \propto\left\{
\begin{aligned}
&\left(\frac{k}{k_{pk}}\right)^{n_{IR}}&(k\ll k_{pk})\\
&\left(\frac{k}{k_{pk}}\right)^{-n_{UV}}&(k\gg k_{pk})
\end{aligned}
\right.
\end{align}
Then the induced GW spectrum is roughly
\begin{align}
    \label{eq:ind_omegagw}
    \Omega_{\GW} \propto\left\{
\begin{aligned}
&\left(\frac{k}{k_{pk}}\right)^{n^{ind}_{IR}}&(k\ll k_{pk})\\
&\left(\frac{k}{k_{pk}}\right)^{-n^{ind}_{UV}}&(k\gg k_{pk})
\end{aligned}
\right.
\end{align}
where
\begin{align}
    n_{IR}^{ind}=\left\{
\begin{aligned}
&2n_{IR}-2b &(n_{IR}<3/2)\\
&3-2|b| &(n_{IR}>3/2)
\end{aligned}
\right.
\end{align}
and
\begin{align}
    \label{eq:eta_uv}
    n_{UV}^{ind}=\left\{
\begin{aligned}
&2n_{UV}+2b &(n_{UV}<4(2))\\
&4(2)+n_{UV}+2b &(n_{IR}>4(2))
\end{aligned}
\right.
\end{align}
where we defined
\begin{align}
b=\frac{1-3w}{1+3w}\,.
\end{align}
The values in parenthesis n Eq.~\eqref{eq:eta_uv} correspond to the case $c_s^2\sim1$ \cite{Balaji:2022dbi}. In the limiting cases of the inequalities,  
as well as in the case of $n_{IR}>3/2$ and $w=1/3$, logarithmic corrections appear. {For the purpose of this discussion, we neglect these effects here.} 

We see that for different values of the parameters $b$, $c_s$, $n_{IR}$ and $n_{UV}$, one can easily obtain different UV and IR scalings of $\Omegagw$ (i.e. $n_{UV}^{ind}$ and $n_{IR}^{ind}$) and mimic the GW signal from the other production mechanisms mentioned above. Thus, in the absence of independent (non-GW) constraints on the scalar power spectrum on small scales, one cannot unambiguously determine the source of the SGWB from the reconstruction of the spectral shape alone.

\bibliographystyle{apsrev4-2}
\bibliography{ref}

\end{document}